
\magnification=\magstep1
\hsize=16.0  true cm
\baselineskip=20pt
\parskip=0.2cm
\parindent=1cm
\raggedbottom

\def\pp{\parshape 2 0truecm 16truecm 1truecm 15truecm}
%
\def\apjref#1;#2;#3;#4 {\par\pp#1,  #2,  #3, #4 \par}
%
%
\def\oldapjref#1;#2;#3;#4 {\par\pp#1, {\it #2}, {\bf #3}, #4. \par}
%
%
\def\apj{ApJ}
\def\aap{A\&A}

\def\mnras{MNRAS}
\def\aj{AJ}
\def\araa{ARA\&A}

\def\apjs{ApJS}
\def\header#1{\bigskip\goodbreak\noindent{\bf #1}}
%
%
\def\subsection#1{\goodbreak\noindent\underbar{#1}}
\def\subsubsection#1{{\noindent \it #1}}
\def\ltsima{$\; \buildrel < \over \sim \;$}
\def\simlt{\lower.5ex\hbox{\ltsima}}
\def\gtsima{$\; \buildrel > \over \sim \;$}
\def\simgt{\lower.5ex\hbox{\gtsima}}
%
%



\def\kms{{\rm\,km\,s^{-1}}}

\newcount\notenumber
\notenumber=1
\def\note#1{\footnote{$^{\the\notenumber}$}{#1}\global\advance\notenumber by 1}

\newcount\eqnumber
\eqnumber=1
\def\new{{\rm(\chaphead\the\eqnumber}\global\advance\eqnumber by 1}
\def\ref#1{\advance\eqnumber by -#1 (\chaphead\the\eqnumber
     \advance\eqnumber by #1 }
\def\last{\advance\eqnumber by -1 {\rm(\chaphead\the\eqnumber}\advance
     \eqnumber by 1}
\def\eq#1{\advance\eqnumber by -#1 equation (\chaphead\the\eqnumber
     \advance\eqnumber by #1}
\def\eqnam#1#2{\immediate\write1{\xdef\ #2{(\chaphead\the\eqnumber}}
    \xdef#1{(\chaphead\the\eqnumber}}

\newcount\fignumber
\fignumber=1
\def\nfig{\the\fignumber\ \global\advance\fignumber by 1}
\def\nfiga#1{\the\fignumber{#1}\global\advance\fignumber by 1}
\def\rfig#1{\advance\fignumber by -#1 \the\fignumber \advance\fignumber by #1}
\def\fignam#1#2{\immediate\write1{\xdef\
#2{\the\fignumber}}\xdef#1{\the\fignumber}}

\newbox\abstr
\def\abstract#1{\setbox\abstr=\vbox{\hsize 5.0truein{\par\noindent#1}}
    \centerline{ABSTRACT} \vskip12pt \hbox to \hsize{\hfill\box\abstr\hfill}}

\def\deg{^\circ}


\def\s{\ifmmode \widetilde \else \~\fi}
\def\={\overline}

\def\spose#1{\hbox to 0pt{#1\hss}}

\def\etal{{\it et al.\ }}
\def\cf{{\it cf.\ }}
\def\eg{{\it e.g.\ }}

\def\lta{\mathrel{\spose{\lower 3pt\hbox{$\mathchar"218$}}
     \raise 2.0pt\hbox{$\mathchar"13C$}}}
\def\gta{\mathrel{\spose{\lower 3pt\hbox{$\mathchar"218$}}
     \raise 2.0pt\hbox{$\mathchar"13E$}}}
\def\Dt{\spose{\raise 1.5ex\hbox{\hskip3pt$\mathchar"201$}}}	
\def\dt{\spose{\raise 1.0ex\hbox{\hskip2pt$\mathchar"201$}}}	


\def\=={\equiv}

\def\dotsfill{\leaders\hbox to 1em{\hss.\hss}\hfill}









\rm		
\def\Atoday{\ifcase\month\or
  January\or February\or March\or April\or May\or June\or
  July\or August\or September\or October\or November\or December\fi
  \space\number\day, \number\year}
\def\Etoday{\number\day\space\ifcase\month\or
  January\or February\or March\or April\or May\or June\or
  July\or August\or September\or October\or November\or December\fi
  \space\number\year}

\def\cl{\centerline}
\def\pasphead#1{\bigskip\goodbreak\cl{\bf #1}\bigskip}
\def\apjl{ApJL}
\centerline {{\bf GALACTIC STRUCTURE AND STELLAR POPULATIONS}}
\bigskip
\cl{ROSEMARY F.G.~WYSE}
\bigskip
\cl{Center for Particle Astrophysics }
\medskip
\cl{University of California, Berkeley, CA 94720}
\cl{and}
\cl {Department of Physics and Astronomy\footnote{$^1$}{Permanent address}}
\cl {Johns Hopkins University, Baltimore, MD 21218}
\bigskip
\cl {Electronic mail: wyse@physics.berkeley.edu}
\bigskip\noindent
Running Head : Galactic Structure
\bigskip
\noindent {\bf ABSTRACT.} The Milky Way Galaxy offers a unique opportunity for
testing theories of galaxy formation and evolution.  I discuss
how large surveys,
both photometric and spectroscopic,
of
Galactic stars are the keystones of
 investigations into such fundamental problems as the merging
history  -- and future --
of the  Galaxy.

\pasphead{1. INTRODUCTION}

The study of the spatial distribution, kinematics and chemical
abundances of stars in the Milky Way Galaxy constrains models of
disk galaxy formation and evolution.  One can address
specific questions such as
\item {(i)} When was the Galaxy assembled?  Is this an ongoing
process? What was the merging history of the Milky Way?
\item {(ii)} When did star formation occur in what is now `The
Milky Way Galaxy'?  Where did the star formation occur then?
What was the stellar Initial Mass Function?
\item {(iii)} How much dissipation of energy was there before and
during the formation of the different stellar components of the
Galaxy?
\item {(iv)} What are the relationships among  the different
stellar components of the Galaxy?
\item {(v)} Was angular momentum conserved during
formation of the disk(s) of the Galaxy?
\item {(vi)} What is the shape of the dark halo?
\item {(vii)} Is there dissipative (disk) dark matter?

These questions are part of larger questions such as
\item {(a)} What was the star-formation rate at high redshift? --
in this context remember that the oldest stars in the Galaxy formed at a
lookback time of 10--15 Gyr, or redshifts significantly greater
than unity.
\item {(b)} Does light trace mass?
\item {(c)} What determines the Hubble Sequence of galaxies?
\item {(d)} What was the power spectrum of primordial density fluctuations?
\item {(e)} What are the values of the cosmological parameters
H$_0$ and $\Omega_0$? For example, the length-scale of the Galactic disk
compared with those of external galaxies constrains H$_0$, on the
presumption that the Milky Way is a typical spiral (van der Kruit
1986).

In this talk I will focus on two (related) questions for which
large surveys of stars in the Galaxy are crucial, namely
`what is the shape of the stellar halo?' and `what is the merging
history of the Milky Way Galaxy?'  The shape of the stellar halo
constrains the shape of the dark halo, especially when the
spatial data are combined with stellar kinematic data, in addition to
containing information about the important physical
processes during the formation of the stellar halo -- was there
large-scale dissipation of  energy during a collapse phase?  Are
there subcomponents to the halo that may indicate  accretion and merging?
The thick disk of the Galaxy may have been formed by a merger
event with a satellite of perhaps the mass of the Large
Magellanic Cloud; the stars of such accreted satellites may
still be identifiable in `moving groups'.  When will the next merger
event occur? Large multi-color databases such
as those soon to be available from the Sloan Digital Sky Survey
and the APS facility in Minnesota will provide necessary input to
these problems.

\pasphead{2. GALAXY FORMATION }

I shall briefly summarise current ideas of galaxy formation and
evolution to place Galactic work in context (see \eg\ Silk and Wyse
1993 for a more complete review).
The nature of Dark Matter determines the way in which structure
forms in the Universe.  The popular theory of Cold Dark Matter
(CDM; \eg\ Blumenthal \etal\ 1984) predicts that the first objects to
collapse under self-gravity are a small fraction of the mass of a
typical galaxy, so that galaxies form by clustering and merging
of these smaller objects.  All density fluctuations are initially
just gas and dark stuff, and the rate at which stars form is very
model-dependent, but feedback from stars plays a major role in the
energy balance.  The rate of merging and growth of mass of a
protogalaxy can be estimated by studying the dark matter, which
is assumed to be dissipationless and hence to have simpler physics than the
baryonic component, but most probably underestimates the
longevity of individual baryonic structures which would be
observed as distinct galaxies (\eg\ Lacey and Cole 1993;
Kauffmann and White 1993; Navarro, Frenk and White 1994).
Lacey and Cole (1993) have a
particularly vivid schematic representation of the merging
process (their Figure 6) as a tree, where time increases from top
to bottom, and the width of the
branches indicates the mass of a particular halo associated with
galactic
substructure. The final
product, the galaxy, is represented by the
single trunk at the ground, and the relative thicknesses of
branches meeting reflect the mass ratios in a merger.
The merging history of a galaxy is then described by the shape
of the tree. The extremes of morphological type may perhaps be
the result of merging histories that are described by the two types of
trees that I at least was taught to draw as a child -- either one main
trunk from top to bottom, with many small branches joining the
trunk at all heights, as a fir tree, or a main trunk that splits
into two repeatedly, more like an English oak tree.
The latter, dominated by
`major mergers' or equal mass mergers, may lead to an elliptical
galaxy.
The former, where
the merging history is dominated by `minor mergers', or very
unequal mass mergers with  a well-defined central core at all
times, may lead to a disk galaxy.
This picture of disk galaxy formation -- building up by accretion
of substructure  onto a central core -- provides a
synthesis of elements of the
much-discussed and previously
apparently mutually exclusive `monolithic collapse' paradigm of Eggen,
Lynden-Bell and Sandage (1962) and the `chaotic' halo formation
envisaged by Searle and Zinn (1978).  One of the reasons for the
popularity of CDM is its analytic tractibility; Gunn (1987)
presents an elegant exposition of the formation of the Milky Way.

Constraining the merging history of the Milky Way is of obvious
importance, as is determination of the star formation history.
Aspects of these problems are briefly discussed here.
\vfill\eject

\cl{\bf 3. IS THE STELLAR HALO ROUND, FLAT, BOTH, }

\cl{\bf OR MORE COMPLEX?}

Dissipationless N-body simulations of hierarchical clustering
tend to form flattened, triaxial halos, with mean axial ratios of
$<b/a> \sim 0.64$ and $<c/a> \sim 0.45$ at 50kpc for a standard CDM
power spectrum (Dubinski and
Carlberg 1991).  Iso-potential surfaces are in general rounder
than  iso-density surfaces -- for reasonable
axisymmetric potentials at a few core radii the ellipticity of the
potential contours is only about one-third that of the density
contours (Binney and Tremaine 1987) -- and, depending on kinematics, one may
expect a  system of
test particles to have an axial ratio that is closer to that
of the potential rather than that of the gravitating
density (for example Binney and May 1986).
Thus one might
expect luminous galaxies to be rounder than their dark haloes.
Further, the effect of the baryonic
component, once it dissipates and
settles to the center of the potential well,  on the dark halo
will be to decrease its flattening and make it more oblate
(Katz and Gunn 1991; Dubinski 1994). However, one
would not expect any  component to
be spherical.

 Counts  of stars in different lines-of-sight can be used
fairly directly to obtain an estimate of the shape of the stellar
halo, although a dependency on models enters through the need to
isolate {\it halo\/} stars; this is generally attempted by
concentrating on blue stars, $B-V \simlt 0.6$.  Analyses
of Schmidt plates to $V
\sim 20$, or distances of unevolved halo stars from the Sun of $\sim 5-6$kpc,
 favor  a flattened stellar halo, $c/a
\sim 0.6$ (Wyse and Gilmore 1989; Larsen and Humphreys 1994).
Deeper star counts, to V$ \sim 21$,
have favored rounder isophotes, $c/a \sim 0.8$
(Bahcall and Soneira 1980; Reid and Majewski 1993).
Accurate star-galaxy-qso separation is obviously crucial
especially for the analysis of faint blue stars.
However, it may
be that there are indeed two components to the stellar halo, with
the more distant stars being in a rounder component.
Kinman, Wirtanen and Janes (1966) found a
general increase in the flattening of the field RR Lyrae
population with increasing Galactocentric distance.
Hartwick
(1987) found that the metal-poor, [Fe/H]$ \simlt -1$ dex,
halo globular clusters and the
metal-poor RR Lyrae stars both had a spatial distribution that
was better fit by two components; his inner component, which dominates
in the solar neighborhood, is flattened with
an axial ratio of $\sim 0.6$, while the outer component, dominant at
$R \simgt 15$kpc is spherical.
Kinman, Suntzeff and Kraft (1994) also found evidence for two
components, one significantly flattened which dominates locally
and one more spherical,
in their sample of  halo blue horizontal branch stars.  However,
these analyses of horizontal branch stars are
hampered by the small number of stars available; studies of unevolved tracers
are not in principle.

The  kinematics of the stellar halo provide only a weak
constraint on the shape at present, but it is difficult to fit
both the local and the distant data simultaneously (Arnold 1990;
van der Marel 1991).  Answering the question of whether or
not the flattened component to the halo
horizontal branch stars is rotationally supported requires more data.

The existence of
two components may be evidence that the stellar halo formed by a
hybrid chaotic/monolithic collapse process as described in section 2 above
(\cf\ van den Bergh
1993; Zinn 1993; Norris 1994), or that  the stellar halo is
 locally flattened in  response to the
disk potential (\cf\ Binney and May 1986),
or perhaps reflects the different orbital parameters and internal
structure of disrupted satellite galaxies that were accreted by
the Galaxy to form the stellar halo (Freeman 1987).  Satellite
galaxies are slowed in their orbit by dynamical friction against
the background  halo (assuming their orbit penetrates the
mass distribution of the Galaxy), depositing  orbital energy and angular
momentum.  The satellite galaxy will be tidally shredded as it
sinks towards the Galactic center, essentially
losing material with density less than the average density of the
background Galaxy interior to the satellite's current position.
Thus the internal structure of the satellite plus its initial
orbit determine how much of the satellite's vertical motion is
damped prior to significant disruption of the satellite. This is
discussed further below.

Thus improved  distant samples are required for halo  kinematics
and also for chemical abundances to constrain the interpretation
of the star count analyses, by better isolation of `halo' stars.
Probably only spectroscopy will allow quantification  of quasar
contamination of representative blue star samples (as will be
carried out by the Sloan Digital Sky Survey), but multi-color
photometry combined with measuring machines such as the APS will
suffice  for the bulk of the stars.
\bigskip

\cl{\bf 4. THE THICK DISK}

\cl {\bf AND THE MERGING HISTORY OF THE MILKY WAY}

\header {4.1 The Thick Disk}

As discussed briefly above, mergers between galaxies and/or
galactic substructure are
inherent in hierarchical-clustering scenarios for the growth of
structure in the Universe. Since the pioneering work of Toomre
(\eg\ 1977) much effort has gone into study of the
effects of both stellar and gaseous mergers upon galaxy
morphology.  The thick disk of the Milky Way Galaxy at least
morphologically could be a minor-merger remnant; provided all the
orbital energy of an accreted satellite, mass $M_{sat}$,
is available to increase
the random energies of the stars in a thin disk of mass
$M_{disk}$, then after a merger the thin disk will be heated by
an amount
(Ostriker 1990) $$\Delta v_{random}^2 = v_{orbit}^2 M_{sat}/M_{disk}.$$
Of course  the internal degrees of freedom of
the satellite could also be excited and any gas present could,
after being heated,
cool by radiation,  so this is a definite upper
limit to the heating of the disk.  This estimate is suggestive, however,
that the thick disk,
which has a vertical velocity dispersion of $\sim 45 \kms$, could
be formed from a thin disk with vertical dispersion of $\sim 20
\kms$ by accretion of a satellite of about 10\% of the mass of
the disk.  Note although there is indeed an age--velocity dispersion
relationship for stars in the thin disk, the value of the vertical
velocity dispersion is observed to saturate at $\sim 20 \kms$
for stars older than a few Gyr (\eg Freeman 1991), which may be
understood if the
heating mechanism responsible is confined to the thin disk itself,
such as scattering by encounters with giant molecular clouds (\eg
reviewed by Lacey 1991).

Mergers between disks and satellites
have been studied by Quinn, Hernquist and
collaborators.  The most detailed N-body simulation they have
published to date (Quinn, Hernquist and Fullager 1993) follows
the evolution of a satellite galaxy, initially in a prograde, circular
orbit about a disk of mass ten times that of the satellite,
with the orbit inclined 30$\deg$ to the plane of the disk, and
at a Galactocentric radial distance of
six disk scalelengths.  Further parameters
that need to be specified define  the internal density profile and
kinematics of the satellite, and of the disk. The generic
evolution is that the satellite's vertical motion is damped
rather quickly, with  a slower radial  decay to the central
regions (displayed very clearly in their Figure 11). The orbital
energy is indeed deposited both in the disk particles and in the
internal degrees of freedom of the satellite;
the final galaxy has a thick disk
 which consists of both heated thin-disk stars and
shredded-satellite stars.  The mix of these obviously depends on
the (many) model parameters, both orbital and internal to the
galaxies.  Both the radial and vertical structures
of the disk are altered. The radial heating and re-arrangement of
material by angular momentum transport  results in an asymmetric
drift (lag behind local circular velocity) that is only $\sim 10
\kms$ higher than the initial disk.  The vertical heating
increases the disk scale-height by about a factor of two.

How does this compare to the observed thick disk in the Milky
Way?  The scaleheight is in reasonable agreement with star count analyses.
Determination of thick disk kinematics is complicated by
the fact that there is strong overlap with the other stellar
populations of the Galaxy, and in particular the thin disk.  The
most recent determinations have used robust decompositions of the
observational data and find that the thick disk asymmetric drift
is constant with height above the plane, at least over 500pc --
2kpc, and has a value of $ V_{lag} \sim 50-60 \kms$ (Soubiran 1993;
Ojha \etal\ 1994), which is not in wonderful  agreement with this particular
simulation.
Of course the `standard' merger of Quinn \etal\ is only illustrative.

The chemical abundance distribution of the thick disk is an obvious
discriminant of models of thick disk formation.  For example, in
the dissipationless stellar accretion scenario, the thick disk
should be a mix of the pre-existing thin disk, and the accreted
satellite.  An alternative picture that can produce sufficient heating to
explain the thick disk appeals to rare close encounters
of thin-disk stars with massive black
holes in the halo, and predicts
that the thick disk should contain the same range of
chemical abundances as the present thin disk
(Lacey and Ostriker
1985).
The available estimates of age for the bulk of the
thick disk imply that its stars
formed early in the evolution of the Galaxy,
perhaps at the same time as the metal-rich globular clusters
(Gilmore, Wyse and Kuijken 1989; Edvardsson \etal\ 1993), which
argues against an on-going process of  disk
thickening.\footnote{$^\heartsuit$}{In fairness to
Lacey and Ostriker they were concerned more with explaining
the population of high-velocity A
stars (\eg\ Lance 1988)
as being  the product of rare close encounters of disk stars
with massive halo
black holes; however  one might also note that massive black holes have
fallen out of favor as candidates for halo dark matter, due to recent
calculations concluding that for masses of $\simgt 10^3 M_\odot$
they are too efficient at heating and would
disrupt halo globular clusters (Moore 1993) and  the
disks of low-mass galaxies (Rix and Lake 1993).}
This also
implies that in the merger scenario
the chemical abundance distribution of the thick
disk should reflect that of the thin disk and satellite perhaps
14~Gyr ago.  While this is rather model-dependent, studies of the
older stars in the present satellites of the Milky Way  might
lead one to expect a metal-poor component to the thick disk, with
metallicities $\simlt -1$ dex, in addition to the contribution
from direct heating of the thin disk, with [Fe/H] $ \simgt -1$
dex.  A metal-poor component of the
thick disk may also be expected if the thick
disk/thin disk formed with no pre-enrichment (the ejecta from
stellar evolution in the stellar halo is, due to its low
angular-momentum per unit mass,
 destined to flow to the central regions of the Galaxy).

 The thick disk is observed to have a  well-defined
peak chemical abundance of $\sim -0.6$ dex (Gilmore and Wyse
1985; Carney, Latham and Laird 1989;  Gilmore, Wyse and Jones
1994) and a fairly broad spread, $\sigma_{[Fe/H]} \sim 0.3$ dex.
Metal-poor stars with thick-disk kinematics have been reported by
Norris, Bessell and Pickles (1987) and by Morrison, Freeman and
Flynn (1991) on the basis of DDO photometry; however as these
stars were discovered as part of a survey of metal-poor stars,
their normalisation relative to the main thick disk is very
uncertain, and may be consistent with determinations of the
metallicity distribution from unbiased surveys with spectroscopic
metallicities.  Further, the
calibration of the DDO
photometric metallicity estimator has recently been called into
question for just the metallicity, [Fe/H] $ \sim -1.2$ dex, of
the `metal-poor thick disk' (Twarog and Anthony-Twarog 1994).
However, the situation is not settled since another survey for
metal-poor stars has found similar results to Morrison \etal\ but
with improved metallicity determinations (Beers and Sommer-Larsen
1994).  Clearly an unbiased, large dataset is required.

\header {4.2 How Much Merging is On-going?}

Do we have a complete census of the satellite galaxies of the
Milky Way?  All-sky surveys, with good photometry in several
bands, and robust star-galaxy separation,
offer the best way to identify overdense regions in
color-magnitude space, and to find clumps of rare stars, for
example carbon stars, which may indicate a physical entity.
The hierarchical clustering and merging picture of galaxy formation
predicts that
there should be many shredded satellite galaxies for every parent galaxy.
Lynden-Bell (1982) noted that the major
axes of the stellar isophotes
of many
of the dwarf spheroidal companion galaxies to the Milky Way
align with
their orbits, as predicted for galaxies that are being tidally
distorted.  This effect has been nicely illustrated by
McGlynn (1990)
in his Figure 2c.  Since particles which become
unbound have low positive energies, they thus only slowly drift away from
their distressed parent galaxy.
As discussed in a slightly different context by
Tremaine (1993), a
tidally disrupted object will spread out into a tidal stream of length
determined by the spread in orbital frequency over the size of the object.
After time $t$, the size of the remnant, $s$,  relative to the radius
of the satellite
at which tidal stripping began, $R_{satellite}$, is
$s/R_{satellite} \sim \Omega \, t$, where $\Omega $ is the angular
frequency of a circular orbit
at the galactocentric radius where the satellite was disrupted.
For disruption at 50kpc,
after 1Gyr the satellite is spread out to
only four times its original size.

 Satellite galaxies  at high Galactic latitude may often
be found by simply
plotting stellar surface density contours projected on the sky.
For example, the dwarf galaxy discovered in the Sextans
constellation, latitude $b = 42.3$, by Irwin \etal\ (1990) is
clearly seen in their Figure 1(b), which plots all stellar
objects down to the plate limit ($B_J \sim 22.5$); note the
paucity of objects classified as stars -- the contours plotted
start at 1.5 images/arcmin$^2$ and increase in steps of 0.5
images/arcmin$^2$. The dwarf has an angular extent of a degree or
so.

Dwarf galaxies could be hiding at lower latitudes, where there
is sufficient foreground disk that the over-density of the
satellite is too small a perturbation to be detected by a simple
sum of all stellar images.  An example of this is the Sagittarius
dwarf, recently discovered (Ibata, Gilmore and Irwin 1994),
as part of a survey of the kinematics
of the bulge (Gilmore and Ibata, in prep).  The dwarf initially
manifest itself as a moving group, with very distinct kinematics
from the bulge, and rather localised in angular position.  The
color-magnitude diagram of the field where the moving group is
seen kinematically contains a fairly obvious core-helium-burning
red `clump' (indicative of an intermediage-age population) and
giant branch, which are not detected in those  fields where
the stars have normal bulge kinematics. Isolating stellar images
which have the  color and apparent magnitude of the clump,
and plotting
isodensity contours of the difference between the `moving group'
field and an offset field, revealed the spatially-extended dwarf
galaxy (Ibata, Gilmore and Irwin 1994).  The distance to the
dwarf may be estimated from the clump/HB magnitude, resulting in
a position $\sim 12$ kpc from the Galactic center on the other
side of the Galaxy, $\sim 5$ kpc below the disk plane.  The dwarf
is inferred to be rather massive, similar to the Fornax dwarf galaxy,
from application of the
observed luminosity-metallicity relation of the extant satellite
galaxies (Caldwell \etal\ 1992), and from the red clump/giant population.
At least three globular clusters have kinematics and distances that are
suggestive of their
being physically associated with the dwarf galaxy.

How many of these are there out there?  Digital databases
provide the ideal searching ground to apply optimised filters to
detect moving groups/satellites.

\pasphead{ACKNOWLEDGEMENTS}

The Center for Particle Astrophysics is supported by the NSF.
I thank the Seaver Foundation for support, and
Roberta Humphreys for organising
the special session at the AAS at which this talk was presented.

\pasphead {REFERENCES}

\parindent=0pt
\apjref Arnold, R. 1990;\mnras;244;465

\apjref Bahcall, J.N. and Soneira, R. 1980;\apjs;44;73

\pp Beers, T. and Sommer-Larsen, J. 1994, preprint

\apjref Bergh, S. van der 1993;\apj;411;178
\apjref Binney, J. and May, A. 1986;\mnras;218;743

\pp Binney, J. and Tremaine, S. 1987, Galactic Dynamics (Princeton,
Princeton Univ. Press)

\apjref Blumenthal, G.R., Faber, S.M., Primack, J.R. and Rees, M.J.
1984;Nature;311;517

\apjref Caldwell, N., Armandroff, T.E., Seitzer, P. and Da Costa, G.S.
1992;\aj;103;840
\apjref Carney, B., Latham, D. and Laird, J. 1989;\aj;97;423
\pp Dubinski, J. 1994, preprint

\apjref Dubinski, J. and Carlberg, R.G. 1991;\apj;378;496

\apjref Edvardsson, B., Andersen, J., Gustafsson, B., Lambert, D.L., Nissen,
P. and Tomkin, J. 1993;\aap;275;101

\apjref Eggen, O., Lynden-Bell, D. and Sandage, A. 1962;\apj;136;748

\apjref Freeman, K.C. 1987;\araa;25;603

\pp Freeman, K.C. 1991, in `Dynamics of Disc Galaxies', ed B.~Sundelius
(G\"oteborgs University, G\"oteborg) p15

\apjref Gilmore, G. and Wyse, R.F.G. 1985;\aj;90;2015

\apjref Gilmore, G. and Wyse, R.F.G. 1986;Nature;322;806

\pp  Gilmore, G., Wyse, R.F.G. and Jones, J.B. 1994, AJ submitted

\apjref  Gilmore, G., Wyse, R.F.G. and Kuijken, K. 1989;\araa;27;555

\pp Gunn, J.E. 1987, in `The Galaxy' eds G.~Gilmore and R.~Carswell (Reidel,
Dordrecht) p413

\pp Hartwick, F.D.A. 1987,
in `The Galaxy' eds G.~Gilmore and R.~Carswell (Reidel,
Dordrecht) p281

\pp Ibata, R., Gilmore, G. and Irwin, M. 1994 Nature 370, 194

\apjref Irwin, M., Bunclark, P., Bridgeland, M. and McMahon, R.J.
1990;\mnras;244;16P
\apjref Katz, N. and Gunn, J.E. 1991;\apj;377;365
\apjref Kauffmann, G. and White, S.D.M. 1993;\mnras;261;921
\apjref Kinman, T.B., Wirtanen, C. and Janes, K. 1966;\apjs;13;379
\pp Kinman, T.D., Suntzeff, N.B. and Kraft, R.P.  1994, preprint

\apjref Kruit, P.C. van der 1986;\aap;157;230

\pp Lacey, C.G. 1991, in `Dynamics of Disc Galaxies', ed B.~Sundelius
(G\"oteborgs University, G\"oteborg) p257

\apjref Lacey, C.G. and Cole, S. 1993;\mnras;262;627
\apjref Lacey, C.G. and Ostriker, J.P. 1985;\apj;299;633
\apjref Lance, C.M. 1988;\apj;334;927
\pp  Larsen, J.A. and Humphreys, R.M. 1994, ApJL submitted

\apjref Lynden-Bell, D. 1982;Observatory;102;202

\apjref Marel, R.P. van der 1991;\mnras;248;515
\apjref McGlynn, T.A. 1990;\apj;348;515
\apjref Moore, B. 1993;\apjl;413;L93
\apjref Morrison, H., Flynn C. and Freeman K.C. 1990;\aj;100;1191
\apjref Navarro, J., Frenk, C.S. and White, S.D.M. 1994;\mnras;267;L1
\pp Norris, J. 1994, preprint

\apjref Norris, J., Bessel, M. and Pickles, A. 1985;\apjs;58;463
\apjref Ojha, D.K., Bienaym\'e, O., Robin, A.C. and Mohan, V.
1994;\aap;284;810
\pp Ostriker, J.P. 1990 in `Evolution of the Universe of Galaxies' ed R.~Kron
(ASP, San Francisco) p25

\apjref Quinn, P.J., Hernquist, L. and Fullagher, D.P. 1993;\apj;403;74

\apjref Reid, I.N. and Majewski, S. 1993;\apj;409;635
\apjref Rix, H.-W. and Lake, G. 1993;\apjl;417;L1
\apjref Searle, L. and Zinn, R. 1978;\apj;225;357
\apjref Silk, J. and Wyse, R.F.G. 1993;Phys Rep;231;293
\apjref Soubiran, C. 1993;\aap;274;181
\pp Toomre, A. 1977 in `Evolution
of Galaxies and Stellar Populations' eds B.M.~Tinsley and R.B.~Larson (Yale
University, New Haven) p401

\pp Tremaine, S. 1993, in {`Back to the Galaxy'} eds
F. Verter and S. Holt (AIP, New York) p599

\apjref Twarog, B.A. and Anthony-Twarog, B.J. 1994;\aj;107;1371

\apjref Wyse, R.F.G. and Gilmore, G. 1989;Comments Ap;13;135

\pp Zinn, R. 1993 in {`The Globular Cluster--Galaxy Connection'} eds
G.H.~Smith and \break
J.P.~Brodie (ASP, San Francisco) p38

\bye